\documentstyle[multicol,aps,prl,epsf]{revtex} 
\begin{document} 
 
\title{Dopant Spatial Distributions: Sample Independent Response Function And 
Maximum Entropy Reconstruction} 
\author{D.\ P.\ Chu and M.\ G.\ Dowsett} 
\address{Department of Physics, University of Warwick, Coventry CV4 7AL, UK} 
 
\maketitle 
 
\begin{abstract} 
We demonstrate the use of maximum entropy based deconvolution to reconstruct 
boron spatial distribution from the secondary ion mass spectrometry (SIMS) depth 
profiles on a system of variously spaced boron $\delta$-layers grown in silicon.  
Sample independent response functions are obtained using a new method which 
reduces the danger of incorporating real sample behaviour in the response.  Although 
the original profiles of different primary ion energies appear quite differently, the 
reconstructed distributions agree well with each other.  The depth resolution in the 
reconstructed data is increased significantly and segregation of boron at the near 
surface side of the $\delta$-layers is clearly shown. 
\end{abstract} 
 
\begin{multicols}{2} 
 
The improvement of the depth resolution achieved in sputter profiling in general and 
secondary ion mass spectrometry (SIMS) in particular has continued over the last ten 
years despite occasional predictions about the limit had been reached~\cite{1}.  
Nevertheless, the resolution achieved directly may not be adequate for future 
generations of semiconductor device material, even if ultralow energies~\cite{2,3} or 
large cluster ions~\cite{4} are employed in the primary beam. 
 
The SIMS mass transport effects due to energy deposition and probe beam 
incorporation from the primary beam are well known --- the measured profile is 
broaden and shifted from its true position~\cite{5,6,7}.  Although the use
of low beam 
energies or high mass clusters can greatly alleviate the effects, such mass transport 
effect is intrinsic to the SIMS depth profiling process and can still be observed even 
when the primary ion beam energy is as low as $250 {\rm eV}$~\cite{8} or when 
using SF$_5^+$ ions at $600 {\rm eV}$~\cite{4}.  The true spatial distributions 
remain to be reconstructed especially when there is an abrupt interface or $\delta$-
doping present. 
 
In the concentration range of common practical interest in SIMS, {\it e.g.} the dilute 
limit where dopant concentration $\le 1\%$, there is a strictly proportional relation 
between the signal and the instantaneous surface concentration as well as between the 
primary ion flux density and erosion rate outside the pre-equilibrium region for a single 
matrix~\cite{9}.  Where the essential physics of the analytical process is linear, 
deconvolution is the mathematically correct method to recover and quantify the depth 
profiles~\cite{10}.  The ideal SIMS signal at the depth $z$, $Y_0(z)$, can be 
expressed as a convolution of the true concentration distribution, $C(z)$, with the 
SIMS instrumental response function, $R(z)$, 
\begin{equation} 
Y_0(z)=\int C(z^\prime) R(z-z^\prime) dz^\prime 
\end{equation} 
since primary ion flux must be also proportional to the elapsed time (otherwise depth 
calibration becomes problematic).  Here we define $Y_0(z)$ and $C(z)$ with the same 
dimension of concentration per unit length and $R(z)$ normalised over the depth to 
simplify the equation and ensure that sample mass is conserved.  A slightly more 
sophisticated model $ Y_0(z)=\int C(z^\prime) R(z-z^\prime, z) dz^\prime$ might be 
used if the depth resolution was depth dependent. 
 
It would be a straightforward inverse problem to determine the true spatial distribution 
$C(z)$ if the corresponding ideal SIMS signal $Y_0(z)$ could be measured.  In fact, 
the measured SIMS signal, $Y(z)$, is as usual a combination of the ideal signal and 
associated non-negligible noise component, $Y_N(z)$, 
\begin{equation} 
Y(z) = Y_0(z) + Y_N(z) 
\end{equation} 
There is no obvious way to find $C(z)$ from $Y(z)$.  Various methods have been 
used to obtain $C(z)$~\cite{3}.  Yet, the lack of objective evidence makes it very 
difficult to reconstruct the real features and separate them from the SIMS effect 
justifiably, {\it e.g.}, to distinguish the segregation and diffusion occurring during 
growth at the interface of two different materials from the SIMS atomic mixing.  
Moreover, the peculiar character of the SIMS depth profile data everywhere positive, 
and large dynamic range (may span 10 orders of magnitude overall and 4-6 orders for 
a particular species), requires a very careful and unbiased treatment.  Manipulating 
data arbitrarily, such as placing a subjective penalty on each change in slope or simply 
filtering certain range of frequency components, could seriously distort the final results 
and/or easily lead to unphysical negative values.  We believe that only the features with 
statistical evidence in the original data should be extracted and a empirical 
deconvolution method~\cite{11} based on the maximum entropy (MaxEnt) 
principle~\cite{12,13} fulfils such a requirement. 
 
The success of a MaxEnt deconvolution method relies on the finding of a sample 
independent response function and a suitable noise model.  Neither of them can be 
obtained for the SIMS process with sufficient accuracy from first principle calculations 
because of incomplete knowledge.  Our recent investigation shows that the noise in the 
SIMS depth profiling follows Poissonian statistics universally~\cite{14} and the 
relation between the mean counts, $s_m(z)$, and its corresponding standard deviation, 
$\sigma_s(z)$, is $\sigma_s(z)=s_m(z)^{1/2}$. 
 
The response function should contain only the SIMS related information, {\it i.e.} the 
broadening, the shift and possibly the ion yield, but  must not contain sample 
dependent features.  Ideally, the response function is the transient measured from an 
infinitesimally thin layer, often known as a $\delta$-layer.  However, such a $\delta$-
layer is an abstraction and even if it were not, we have no means to recognise its 
existence, other than very locally.  Moreover, as the intrinsic resolution of SIMS is 
improved by using sub-keV probes~\cite{8}, it is readily apparent that the real 
approximations to such structures which can be grown leave a measurable sample-
related shape content in the SIMS profile due to statistical placement of atoms, 
segregation, and diffusion at the growth temperature, etc.  Simply using such data as 
response function, deconvolution will suppress real features in the depth profile, and 
produce artificial concentration slopes and unrealistically small feature widths. 
 
In the following, we outline a method for extracting a sample independent SIMS 
response function for the case of boron in silicon sampled by oxygen beam for various 
primary beam energies.  Subsequently, we demonstrate a MaxEnt deconvolution to 
reconstruct the dopant distribution from SIMS depth profiles using the corresponding 
response function and noise model.  The results are then discussed. 
 
Experimental and theoretical studies explored the mass transport in the SIMS 
process~\cite{15,16,17,18}.  It has been shown that the normalised
response function can be 
represented by the following form in several orders of magnitude~\cite{18}: 
\begin{equation} 
\begin{array}{lll}
R(z)&=&\displaystyle{1 \over {2(\lambda_g+\lambda_d)}} \times \\[4mm]
{} & {} &
\left\{  \left[ 1- {\rm erf}(\xi_g) \right] {\rm exp}\left[
z/\lambda_g + 
(\sigma/\lambda_g)^2/2 \right] \right. \\[2mm]
{} & {} & + \left. \left[ 1+ {\rm erf}(\xi_d) \right] {\rm exp}\left[
-z/\lambda_d + (\sigma/\lambda_d)^2/2 \right] \right\} 
\end{array} 
\end{equation} 
where $\xi_g=(z/\sigma+\sigma/\lambda_g)^{1/2}$ and $\xi_d=(z/\sigma-
\sigma/\lambda_d)^{1/2}$, $\sigma$ is the primitive standard deviation, $\lambda_g$ 
and $\lambda_d$ the growth and decay lengths.  The smaller the $\lambda_g$ and 
$\lambda_d$ are, the sharper the $R(z)$ will be.  When they approach zero, the 
$R(z)$ will degenerate to a Gaussian distribution with the deviation $\sigma$.  All 
these parameters of the response function apparently depend on primary ion beam 
energy, $E_p$, and should be monotonically increasing functions of it.  If we could 
find a perfect $\delta$-layer, we might be able to fit the measured data with $R(z)$ to 
obtain these parameters for a certain $E_p$ and use them directly to deconvolve other 
SIMS profiles.  Therefore, the difficulty here is how to determine the $\sigma$, 
$\lambda_g$ and $\lambda_d$ from the measured data justifiably, since there are no 
other techniques which we can use to check the results.  We have to substantiate our 
choices through statistical and trend analysis. 
 
Before we study the $R(z)$ of a specimen in a crystalline substrate, we first look at the 
$R(z)$ of the same specimen in a corresponding amorphous substrate.  This is because 
that the amorphous $\delta$-layers are grown at very low temperature where 
segregation is negligible and only broadening due to diffusion is present.  Hence the 
amorphous $R(z)$ should be qualitatively the same as the true crystalline one.  We 
measured an amorphous boron $\delta$-layer in amorphous silicon grown at room 
temperature by molecular beam epitaxy (MBE) with the range of $E_p$ from $335 
{\rm eV}$ to $11 {\rm keV}$ and fitted the measured profiles with the $R(z)$ in 
Eq.~(3).  The primary ions used are normally oxygen incident.  The obtained 
parameters are plotted against $E_p$ in Fig.~1.  Within the error of measurements, we 
have $\sigma^{amph}=0.86+0.27 E_p^{0.82}$, $\lambda_g^{amph}=0$ and 
$\lambda_d^{amph}=1.60 E_p^{0.54}$, where the lengths are in nm and $E_p$ in 
keV.  This reveals that the SIMS mass transport effect itself has no contribution to 
$\lambda_g^{amph}$.  We believe the same is true in crystalline case.  Moreover, as 
the SIMS mass transport effect will be minimised at zero beam energy, it is reasonable 
to think that the $\lambda_d^{amph}$ is only due to the SIMS effect and the residual 
$\sigma^{amph}$ at $E_p=0$ comes entirely from the boron amorphous $\delta$-
layer itself.  This is consistent with the well known fact that a conserved diffusive point 
source has a Gaussian spatial distribution.  Consequently, the SIMS contribution to the 
amorphous primitive deviation, $\sigma^{amph}$, can be obtained from 
$\sigma_s^{amph}(E_p)^2=\sigma^{amph}(E_p)^2-\sigma^{amph}(0)^2$. 
 
Crystalline $R(z)$ is built up by fitting the SIMS data of a MBE grown boron 
$\delta$-layer in a crystalline silicon substrate with the range of $E_p$ from $250 {\rm 
eV}$ to $11 {\rm keV}$ under the same experimental condition as for the amorphous 
study.  Considering that the bonding energy of atoms in a crystalline material is usually 
larger than its amorphous counterpart, we expect that both $\sigma^{cryst}$ and 
$\lambda_d^{cryst}$ are smaller than the amorphous ones and $\lambda_g^{cryst}$ 
should remain zero.  The boron $\delta$-layer structure grown by MBE normally has a 
segregated interface at the near surface side and a almost ideally abrupt interface at the 
other side.  This will enable us to get a reliable $\lambda_d^{cryst}$ from the fitting.  
For $\lambda_g^{cryst}$, the fit shows no significant energy dependence and we 
therefore take it as intrinsic to the sample.  The results are also shown in Fig.~1.  We 
obtain that $\sigma^{cryst}=0.27+0.39 E_p^{0.75}$, and $\lambda_d^{cryst}=1.39 
E_p^{0.56}$.  These parameters are indeed smaller than the amorphous ones, and 
again $\lambda_d^{cryst}$ vanishes as $E_p$ approaches zero.  The $\sigma^{cryst}$ 
can be partly affected by the sample structure in use and, indeed, shows a finite 
intercept.  Therefore, we take the intercept away and calculate the SIMS part, 
$\sigma^{cryst}$, as $\sigma_s^{cryst}(E_p)^2=\sigma^{cryst}(E_p)^2-
\sigma^{cryst}(0)^2$.  

Using the obtained parameters, we are now able to calculate sample independent SIMS 
response functions for various $E_p$ and use them with our noise model to 
deconvolve some measured SIMS depth profiles by MaxEnt method.  A multiple 
boron $\delta$-layer structure in silicon was grown at a constant $450 {\rm C}$ by 
MBE with the arrangement of, in turn from the surface, a pair of $\delta$-layers $2 
{\rm nm}$ apart, then another pair of $\delta$-layers $5 {\rm nm}$ apart, etc.  The 
intended boron concentration was $1\times 10^20 {\rm atoms/cm^3}$ for all the 
layers, which is well above the bulk solid solubility limit~\cite{19,20}.  We profiled the 
sample using a normally incident oxygen beam at $E_p=0.25$, $0.50$, $1.0$, $2.0$, 
$4.0$, and $6.0 {\rm keV}$, respectively.  The depth calibration of profiles in silicon 
using oxygen primary ions has been discussed and clarified recently~\cite{21}.  In our 
case, since the part we study is outside the pre-equilibrium region and much longer 
than the transition depth we can align centroids of the profiles based on the principle 
that distance between centroids will not be changed by any convolution process.  The 
MaxEnt algorithm used in our calculation is from Ref.~\cite{22}.  The measured 
profiles after conventional calibration and corresponding deconvolved spatial 
distributions for the first two pairs of boron $\delta$-layers are shown in Fig.~2(a) and 
(b), respectively. 
 
\begin{figure}[hbt] 
\narrowtext 
\epsfxsize=3in 
\epsfbox{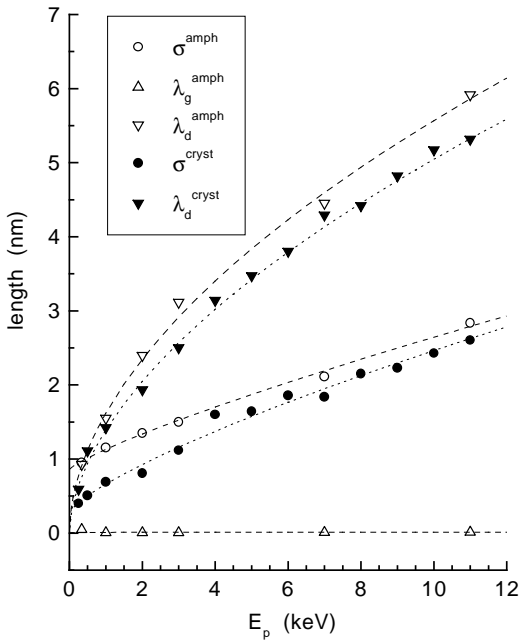} 
\caption{The response function parameters, $\sigma$, $\lambda_g$, and $\lambda_d$ 
vs. primary beam energy $E_p$.  The open and solid symbols as well as the dashed 
and dotted lines are for the amorphous and crystalline parameters, respectively.  The 
lines are fitted with form $a+b~E_p^c$ for each of the parameters.} 
\label{fig1} 
\end{figure} 
 
From Fig.~2(a) and (b), we can see that although the original SIMS profiles appear 
quite differently, the reconstructed spatial distributions agree well with each other.  
The area under each feature is the same before and after deconvolution, {\it i.e.}, the 
number of the boron atoms for each feature is conserved.  It is obvious that the depth 
resolution of the reconstructed spatial distributions has increased significantly.  The 
$\delta$-layers at $z=87 {\rm nm}$ which were separated by $2 {\rm nm}$ can just be 
distinguished at $E_p=2 {\rm keV}$ while the $5 {\rm nm}$ pair at $z=135 {\rm 
nm}$ are well separated at the maximum $E_p=6 {\rm keV}$ we used.  Note that the 
deconvolved features become smoother as $E_p$ increases, {\it i.e.}, better depth 
resolution can be achieved as $E_p$ gets lower.  From the reconstructed dopant 
distribution in Fig.~2(b), it is clearly shown that there is considerable boron 
segregation on the near surface side of the layers.  This kind of feature would be 
automatically eliminated if a simplistic $R(z)$ were used.  There are some unexpected 
small concentration spikes in the reconstructed data at level of two to three orders 
lower than the peak height.  We have no objective criteria to confirm their existence or 
take them away so they remain whether one likes them or not.  
 
\begin{figure}[hbt] 
\narrowtext 
\epsfxsize=3in 
\epsfbox{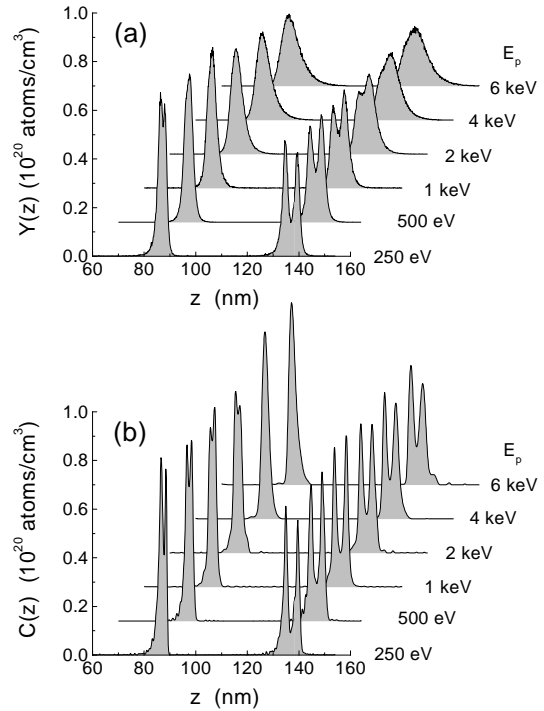} 
\caption{(a) The SIMS depth profiles at different primary beam energy $E_p$ for the 
boron $\delta$-layers in crystalline silicon substrate  and (b) the corresponding spatial 
distributions reconstructed with the SIMS response function through a MaxEnt 
deconvolution.} 
\label{fig2} 
\end{figure} 
 
Using Eqs.~(1) and (2), we can easily work out the corresponding generalised 
Rayleigh limit of depth resolution, $\Delta z$, for two adjacent ideal $\delta$-layers 
after deconvolution if we assume that $\Delta z$ is limited only by the SIMS noise: 
\begin{equation} 
\Delta z(E_p) \gg \displaystyle{2 \over {Y_C^{1/2}}} \lambda_d(E_p) 
\end{equation} 
where we take $\lambda_g=0$, $\lambda_d \gg \sigma_s$, and $Y_C$ is the original 
measured counts.  Estimating with the typical peak counts $Y_C\sim 10^4$ in our 
experiment and the $\lambda_d$ from Fig.~1, we find the resolution we have achieved 
in the above deconvolution is within the limit. 
 
Maximising the entropy of spatial distribution gives us the most likely deconvolved 
solution.  It has a global tendency to spread the solution onto the whole space range 
within a given noise deviation when total concentration is fixed.  However, there is no 
constraint on the local changes of the distribution.  For example, two spikes in the 
distribution contribute the same entropy no matter whether they are next to each other 
or not.  This, unlike local constrains such as limiting the derivative of spatial 
distribution, makes possible the reconstruction of some abrupt features, {\it e.g.} 
$\delta$-layers, superlattice structures, step doping materials, etc. 
 
\begin{figure}[hbt] 
\narrowtext 
\epsfxsize=2.5in 
\epsfbox{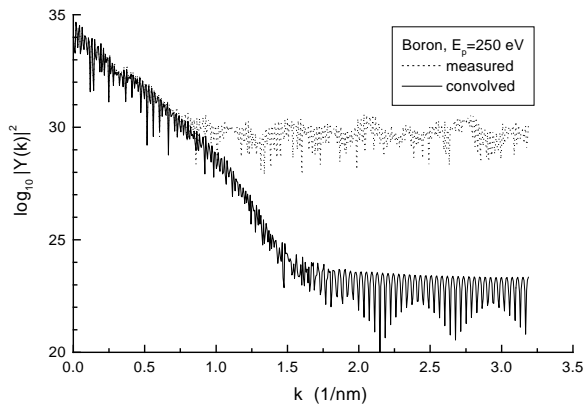} 
\caption{The Fourier power spectrum, $\left| Y(k) \right|^2$, of the measured SIMS 
signal $Y(z)$ vs. wave number, $k$.  The data are calculated from the $0.25 {\rm 
keV}$ SIMS boron profile and the convolved one from the corresponding 
reconstructed distribution.} 
\label{fig3} 
\end{figure} 

To have some further understanding on the MaxEnt deconvolution method, we 
compare the frequency components of the SIMS measured profile and the 
corresponding convolved one calculated from the reconstructed distribution.  Fig.~3 
shows the power spectra for the $E_p=0.25 {\rm keV}$ case.  Although the MaxEnt 
method only manipulates data in real space, the background noise frequency 
components in the convolved profile are clearly suppressed by $6$-$7$ orders.  It is 
clearly shown that the high frequency features in the range of $0.75$ to $1.50 {\rm 
nm^{-1}}$ are retained rather than eliminated if a conventional $1/f$ noise subtraction 
method were used.  Profiles of other $E_p$ have the similar results.  This is consistent 
with Shannon's entropy loss theorem for signal passed through linear filters~\cite{23}.  
Consequently, the MaxEnt method is not only able to reduce the background noise 
level without using any artificial windows in real space or filters in frequency space but 
also capable of retaining sharp features which are statistically significant. 
 
We should emphasis that although the MaxEnt method does improve the depth 
resolution and recovery sharp features, it {\it only} provides us with the statistically 
evident information which the original SIMS profile contains.  The MaxEnt 
deconvolution is not an alternative to instrumental improvement, such as achieving 
lower primary beam energy and higher corresponding beam current.  For example, one 
cannot distinguish whether there are two close $\delta$-layers or a single $\delta$-
layer of higher concentration if the profile is measured with very high beam energies.  
This is because high energy profiling will not only have large $\lambda_d$ but also 
lead to large depth increment and low sampling density, which will limit the 
information in the data at the first place.  To reduce the depth increment for higher 
sampling density in this situation requires significant reductions in the primary beam 
current because of the high sputter yield at high beam energy.  This will lead to lower 
values for $Y_C$.  Therefore, the signal-noise ratio will decrease, which limits the 
potential increase of depth resolution from the deconvolution. 
 
We describe a procedure to obtain sample independent response function which is 
suitable for use where the intrinsic SIMS effect is so small that sample related features 
can be seen.  Using this together with an empirically determined noise model, MaxEnt 
deconvolution is performed to reconstruct self-consistent dopant distributions from the 
SIMS depth profiles obtained at different beam energies.  There are no adjustable 
parameters involved in obtaining the response function and deconvolving the profiles.  
The depth resolution of the reconstructed distributions has been greatly improved and 
the segregation on the near surface sides is clearly demonstrated.  With reconstructed 
true spatial distributions, further quantitative investigations on interface segregation, 
atomic diffusion, and related problems will be straightforward.  Our method can be 
used to study other dopants, and the MaxEnt formalism may be extended to models 
other than convolution. 
 
This work is supported by the EPSRC funding under the grant GR/K32715.  The 
development of the ion column used in this work was funded by the R.~W.~Paul 
Instrument Fund and Atomika GmbH.

\end{multicols}
\end{document}